\documentclass[numbib]{imaiai}
\usepackage{amsfonts,amssymb,amsmath,latexsym,epsfig} 

\newcommand{\pz}{{\mathbb P}}

\begin{document}

\title{Symmetric motifs in random geometric graphs}

\shorttitle{SYMMETRIC MOTIFS IN RANDOM GEOMETRIC GRAPHS} 
\shortauthorlist{C. P. DETTMANN AND G. KNIGHT} 

\author{{
\sc Carl P. Dettmann}$^*$,\\[2pt]
School of Mathematics, University of Bristol, University Walk, Bristol BS8 1TW, UK\\
$^*${\email{Corresponding author: carl.dettmann@bristol.ac.uk}}\\[2pt]
{\sc and}\\[6pt]
{\sc Georgie Knight} \\[2pt]
School of Mathematics, University of Bristol, University Walk, Bristol BS8 1TW, UK\\
{g.knight@bristol.ac.uk}}

\maketitle

\begin{abstract}
{We study symmetric motifs in random geometric graphs. Symmetric motifs are subsets of nodes which have the same adjacencies. These subgraphs are particularly prevalent in random geometric graphs and appear in the Laplacian and adjacency spectrum as sharp, distinct  peaks, a feature often found in real-world networks. We look at the probabilities of their appearance and compare these across parameter space and dimension. We then use the Chen-Stein method to derive the minimum separation distance in random geometric graphs which we apply to study symmetric motifs in both the intensive and thermodynamic limits. In the thermodynamic limit the probability that the closest nodes are symmetric approaches one, whilst in the intensive limit this probability depends upon the dimension. }
{Random geometric graph, spectrum, motif, Chen-Stein method.}
\\
2010 Math Subject Classification: 90B15, 47A10, 05C50, 05C80  	
\end{abstract}

\section{Introduction}
\label{Sec:Intro}

\begin{figure}[ht!]
\begin{center}
\includegraphics[width=6cm]{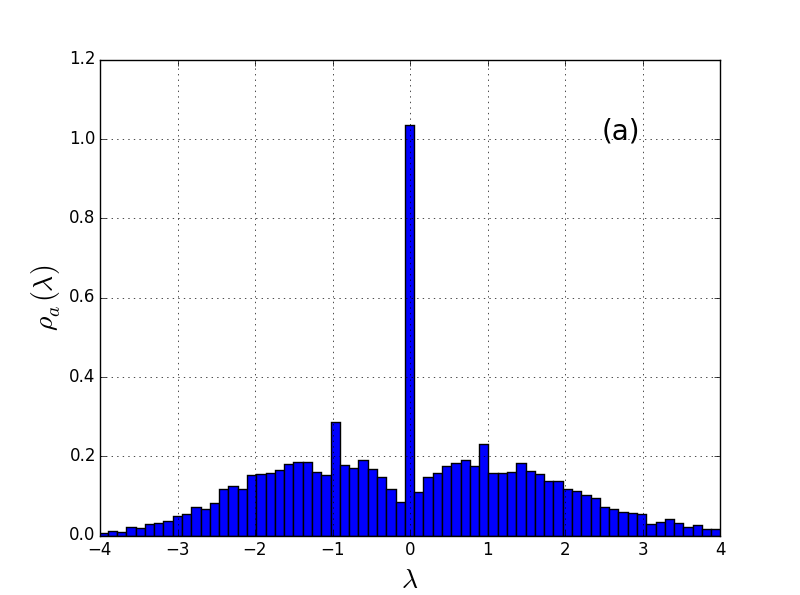} \includegraphics[width=6cm]{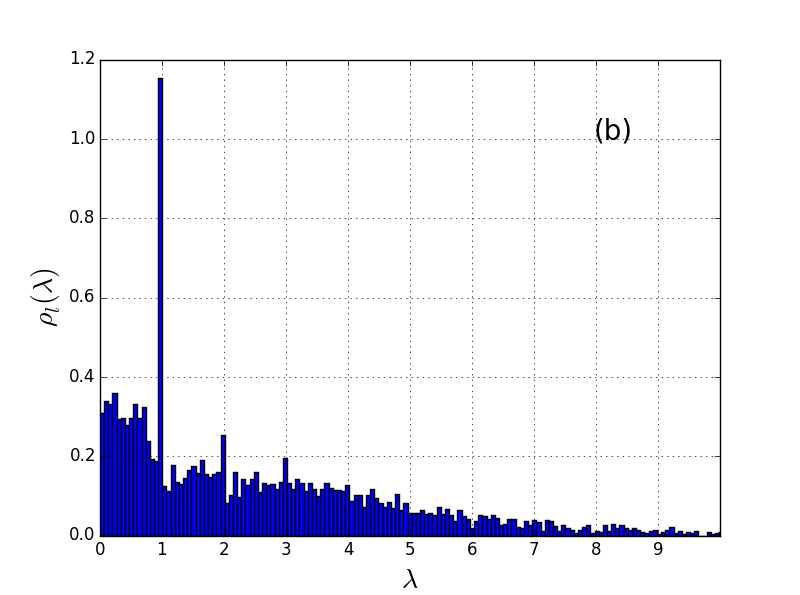}
\caption{Spectral density of the adjacency (a) and Laplacian (b) matrices of the Western States Power Grid of the United States \cite{WaSt98}.  The $4941$ nodes in this network represent the infrastructure of transformers, substations, and generators whilst the edges represent high-voltage transmission connections. Note the peaks at integer values. This network contains $64$ type-I symmetric nodes and $596$ type-II symmetric nodes. The data is available at http://konect.uni-koblenz.de/networks/opsahl-powergrid.}
\label{fig:Power}
\end{center}
\end{figure}

\begin{figure}[ht]
\begin{center}
\includegraphics[width=7cm]{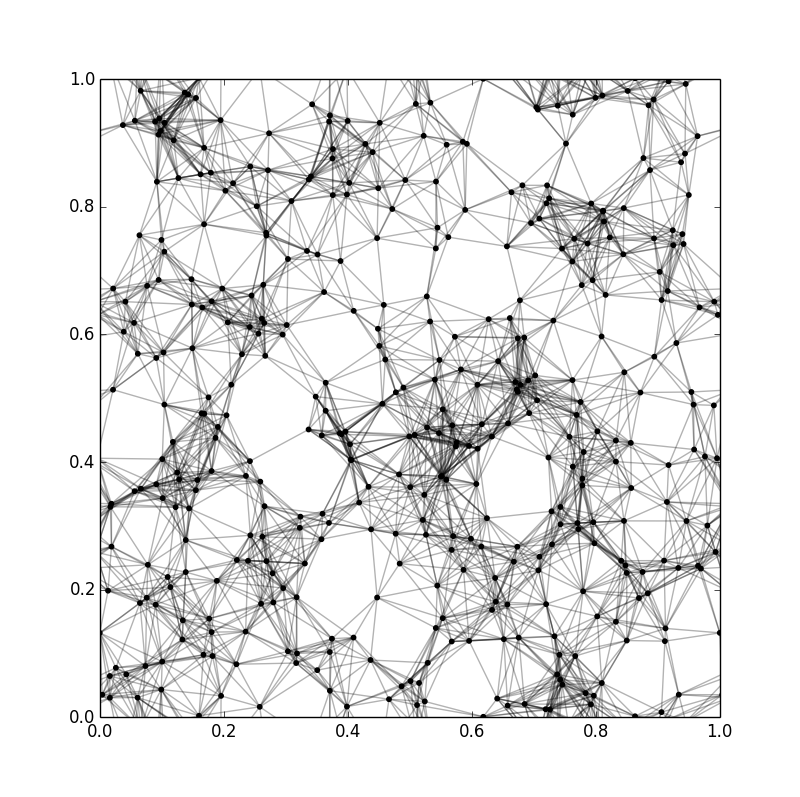}
\caption{Random geometric graph on the torus. An illustration of a $400$ node RGG with a connection range of $r=0.112$. The nodes (black dots) have been distributed randomly, uniformly and are connected by an edge (straight black line) when they are within a Euclidean distance of $r$ with periodic boundary conditions.}
\label{fig:rgg}
\end{center}
\end{figure}

Many physical systems like social networks, biological networks, transport networks and technological infrastructures can be modelled using the graph concept of a set of nodes connected by edges. For an introduction see for example \cite{NewRev, NewBook}. In the study of complex networks, a popular technique is to randomly generate graphs in such a way as to capture certain features of the topology and dynamics of the real-life systems of interest. Some common examples of these are the Erdos-Renyi random graph \cite{ErRen59}, the Barab{\'a}si-Albert scale-free network generator \cite{BarAlb99} and the Watts-Strogatz small-world network generator \cite{WaSt98}. A powerful tool for analysing the topology and dynamics of these networks is their graph spectra \cite{ChungBook} \cite{FarkasEtAl2001} \cite{MieghemBook}. In particular, important subgraphs called symmetric motifs can be seen through the presence of sharp peaks in the spectra indicative of the multiplicity of particular eigenvalues,  see figure \ref{fig:Power} for an illustration of this in the real-world network of the high-voltage power grid in the Western States of the United States of America \cite{WaSt98} and see \cite{MacSanch2008} for an analysis of the symmetric motifs in this particular network. Networks such as this are important for an understanding of efficiency and robustness of power-grids \cite{PhadkeBook}.
It is known that the presence of symmetric motifs is important for other real-world complex systems \cite{MacEtAl2008} and has been shown to influence synchronisation processes \cite{Arenas2006}, \cite{DiazEtAl2009} and dynamical stability \cite{AufEtAl2012}, \cite{DoEtAl2012} and is also related to redundancy and network stability \cite{MacSanch2008}. However, the non-spatial random graph generators mentioned above rarely contain these important subgraphs. It is known, however, that they do occur in a spatial random graph model the random geometric graph (RGG). RGGs were first introduced by Gilbert as a way of modelling wireless networks \cite{Gil61}.  In a RGG the nodes are distributed throughout a given domain uniformly at random and are connected by an edge when they are within a given range of each other. See \cite{PenroseBook}  and \cite{WalSurvey} for introductions and see figure \ref{fig:rgg} for an illustration of a RGG.

RGGs are often used to model networks in which the node location is an important factor, so called spatial networks ( See Ref.\cite{Bart11} for a review). In particular RGGs have been used in modelling wireless networks \cite{GupHum99}, \cite{HaeEtAl2009}, \cite{Pot2000}, \cite{EstEtAl99}, in the study of epidemics \cite{Wang2009},  \cite{Nek2007}, \cite{TorGuc2007}, in the study of city development \cite{Wat2010}, in modelling the vulnerability of infrastructure \cite{XiYe2011} and biological protein-protein interaction networks \cite{HigEtAl2008}. In addition the properties of RGGs such as  synchronisation \cite{EstGua2015},\cite{DiazEtAl2009}, consensus dynamics \cite{EstShe2016}, connectivity \cite{DetOre2016} and spectral properties of RGGs \cite{Nyberg2015}, \cite{BEJ06} have all been studied.

Here we study the symmetric motifs in RGGs. Our aim is to understand how properties like density and dimension of the RGG affect their appearance. We consider the probability of finding symmetric nodes as a function of the connection radius and in two high density limit cases.  In particular, we analyse the symmetric motifs in the {\em intensive limit} of fixed connection radius, and the {\em thermodynamic limit} of fixed mean degree.  It turns out that the expected number of symmetric motifs is non-trivially dependent on these system parameters.  

We will study the binomial model of RGGs. That is, we distribute $N$ points representing $N$ nodes of a network uniformly on the unit torus. A connection is made between two nodes if their toral distance is less than some given range $r$. This is known as the unit disc connection model, but note there are many other connection models in which the links are random with probability depending on the inter-node distance \cite{DetOre2016}. 

In section \ref{sec:Spectrm} we will look at the adjacency and Laplacian spectrum of some RGG ensembles and discuss in particular the sharp peaks we find in these spectra. We will explain how the symmetric motifs which give rise to these sharp peaks and study the probability of finding them in RGGs. In section \ref{sec:dimensions} we will then use the {\em Chen-Stein method} \cite{AGG} to derive the scaling of the minimum separation distance in RGGs. The Chen-Stein method is a tool for obtaining a bound on the total variation distance between a stochastic process which contains dependent random variables and a corresponding independent process. Here, we apply it to remove the dependence of inter-node distances, and hence obtain the variation of the symmetric motifs with dimension in the above limits.  Section \ref{sec:summary} contains a summary.

\section{Spectrum}
\label{sec:Spectrm}
We first look at the spectrum of the graph adjacency matrix $\mathbf{A}$ and Laplacian matrix $\mathcal{L}$. The adjacency matrix is the zero-one adjacency matrix whose entries $a_{ij}=1$ if there is a connection between nodes $i$ and $j$ and zero otherwise. The Laplacian matrix $\mathcal{L}=\mathbf{D}-\mathbf{A}$ where $\mathbf{D}$ has entries $d_{ij}=k_i\delta_{ij}$, $k_i$ the degree of vertex $i$. Again see figure \ref{fig:Power} for an illustration of the spectrum of $\mathbf{A}$ and $\mathcal{L}$ in the real-world network of the high-voltage power grid in the Western States of the United States of America \cite{WaSt98}. 

\begin{figure}[ht!]
\begin{center}
\includegraphics[width=6cm]{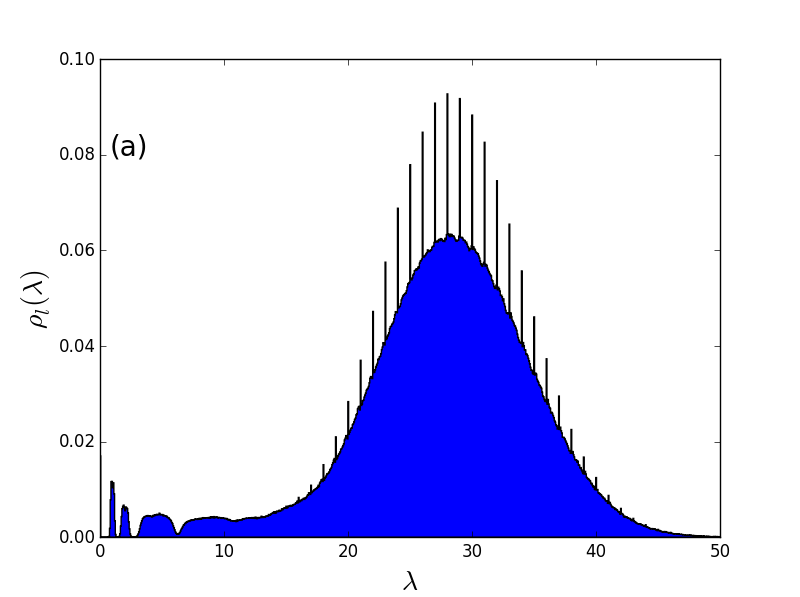} \includegraphics[width=6cm]{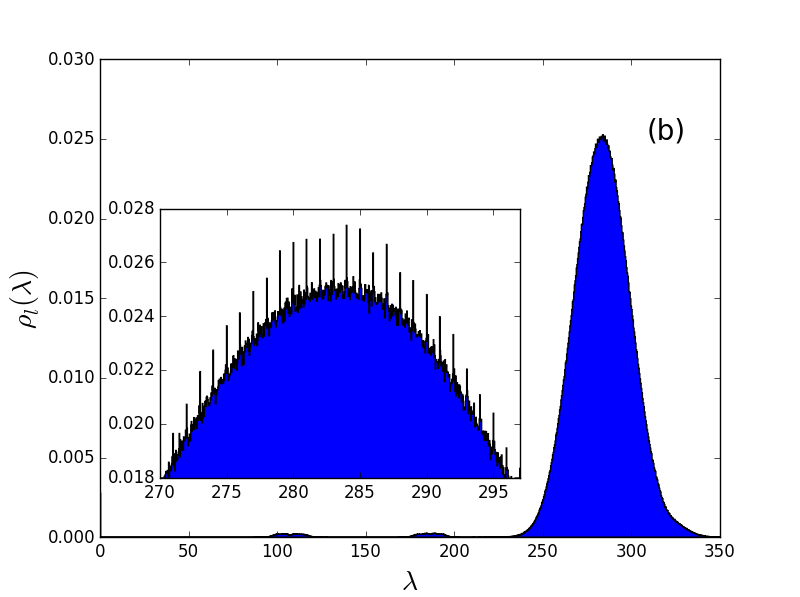}
\caption{Ensemble-averaged Laplacian spectral density. An illustration of the ensemble-averaged spectral density of $10^3$ node RGGs with a connection range of $r=0.09375$ (a) and $r=0.3$ (b). The ensembles consist of $10^4$ RGGs. Inset in (b) shows detail of peaks at integer values.}
\label{fig:lspec}
\end{center}
\end{figure}

\begin{figure}[ht!]
\begin{center}
\includegraphics[width=6cm]{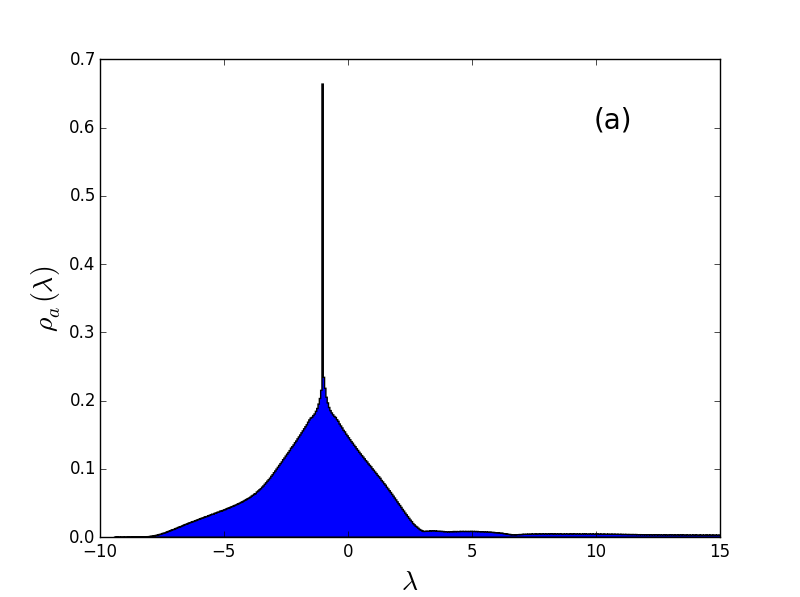} \includegraphics[width=6cm]{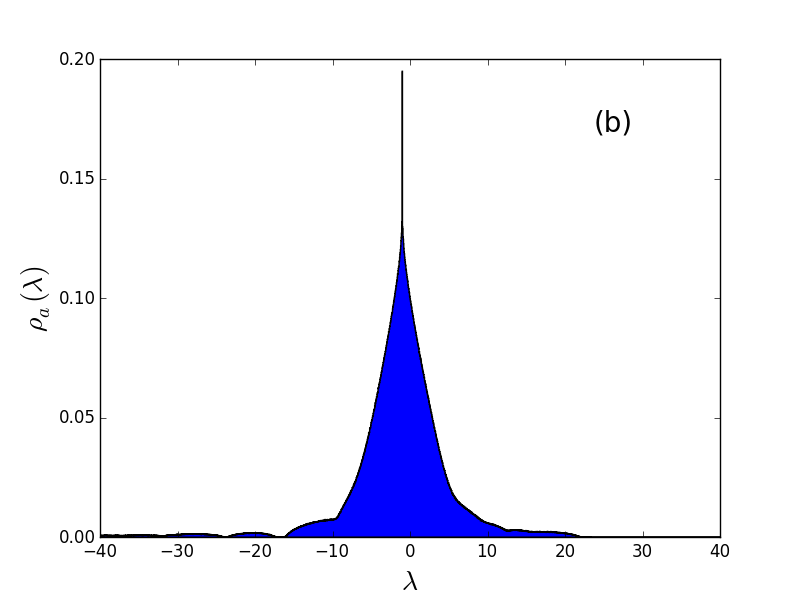}
\caption{Ensemble-averaged adjacency spectral density. An illustration of the ensemble-averaged spectral density of $10^3$ node RGGs with a connection range of $r=0.09375$ (a) and $r=0.3$ (b). The ensembles consist of $10^4$ RGGs.}
\label{fig:aspec}
\end{center}
\end{figure}

In \cite{Nyberg2015} the Laplacian spectra of one-dimensional RGGs  is studied. The authors show that the ensemble-averaged Laplacian spectrum of RGGs on the circle consist of a continuous part and a discrete part consisting of peaks at integer values. We numerically obtained the Laplacian spectrum for an ensemble of RGGs. The ensemble-averaged spectral density $\rho_l(\lambda)$ is illustrated in figure \ref{fig:lspec}. In \cite{BEJ06} the spectra of the adjacency matrix is studied for RGGs. They find that the ensemble averaged spectral density has a discrete part consisting of a peak at $-1$. We numerically obtained the adjacency spectrum of an ensemble of RGGs. This is illustrated in figure \ref{fig:aspec}.


In both \cite{Nyberg2015} and \cite{BEJ06} they identify the presence of symmetric nodes as the structural phenomenon which gives rise to the multiplicities in the eigenvalues which characterise the spectral densities. These {\em Motifs} or {\em graph orbits} are subgraphs whose nodes are invariant under permutation of the indices. That is two nodes are symmetric in this sense when they are connected to the same set of nodes. Eigenvectors localise on these symmetric nodes and give rise to integer eigenvalues. To see this let $n_1$ and $n_2$ be symmetric nodes and $\mathbf{x}$ be a vector with $x_1=1$ and $x_2=-1$, all other entries equal to zero. Considering the adjacency matrix we then have

\begin{equation}
		\left( \begin{array}{cc|c}
0&1&...\\
1&0&...\\
\hline
1&1&\\
0&0&\\
:&:&
\end{array} \right)\left(  \begin{array}{c}
1\\
-1\\
0\\
0\\
:
\end{array} \right)
=-1\left(  \begin{array}{c}
1\\
-1\\
0\\
0\\
:
\end{array} \right)
\label{Eq:motifadj}
\end{equation}
if  $n_1$ and $n_2$  are connected (called a Type-I symmetry). If they are not connected we get eigenvalue $0$ (a Type-II symmetry). For the Laplacian we get

\begin{equation}
		\left( \begin{array}{cc|c}
k&-1&...\\
-1&k&...\\
\hline
-1&-1&\\
0&0&\\
:&:&
\end{array} \right)\left(  \begin{array}{c}
1\\
-1\\
0\\
0\\
:
\end{array} \right)
=(k+1)\left(  \begin{array}{c}
1\\
-1\\
0\\
0\\
:
\end{array} \right)
\label{Eq:motiflap}
\end{equation}
if  $n_1$ and $n_2$  are connected. If they are not connected we get eigenvalue $k$. Here $k$ is the degree of the vertices $n_i$. We note that the Laplacian distinguishes between different degrees in these motifs whilst the adjacency does not. Note that for a set of $s$ symmetric nodes, there are $s-1$ independent, orthogonal eigenvectors that lead to a multiplicity of $s-1$ for the eigenvalue.

\subsection{Symmetry probability}
\label{subsec:symm}

The symmetry of a node is dependent upon the nodes in its neighbourhood and their respective neighbourhoods. The neighbourhood of a node $n_i$ is defined as $B_r(n_i)= \{ x : |x - n_i| \leq r\}$, that  is the region within the connection range of $n_i$. Two nodes have a shared neighbourhood $\mathcal{N}_{s}(n_i,n_j) = B_r(n_i) \cap  B_r(n_j)$ which is given by the intersection of their respective neighbourhoods. An additional important concept is the excluded neighbourhood, $ \mathcal{N}_{ex}(n_i, n_j) = (B_r(n_i)\cup  B_r(n_j)) \backslash (B_r(n_i) \cap  B_r(n_j)) $ which for two vertices is the region that is within the range of one of the vertices but not both. 

A given node $n_i$  is type-I symmetric when there is  a node $n_j$ within its neighbourhood and when the excluded neighbourhood of $n_i$ and $n_j$ is empty. This ensures that $n_i$ and $n_j$ are connected and that they are connected to the same set of nodes. If they are not within the  neighbourhood of each other but have an empty excluded neighbourhood then they are type-II symmetric.

\begin{figure}[htb]
\begin{center}
\includegraphics[width=6cm]{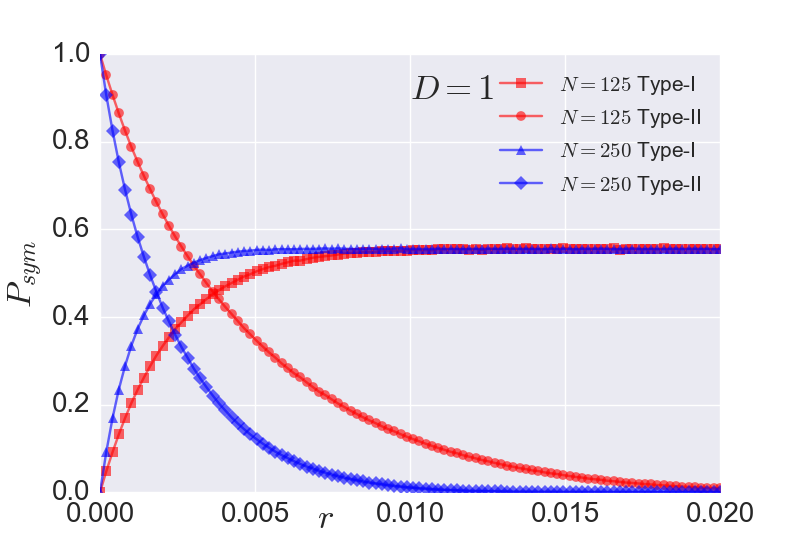} \includegraphics[width=6cm]{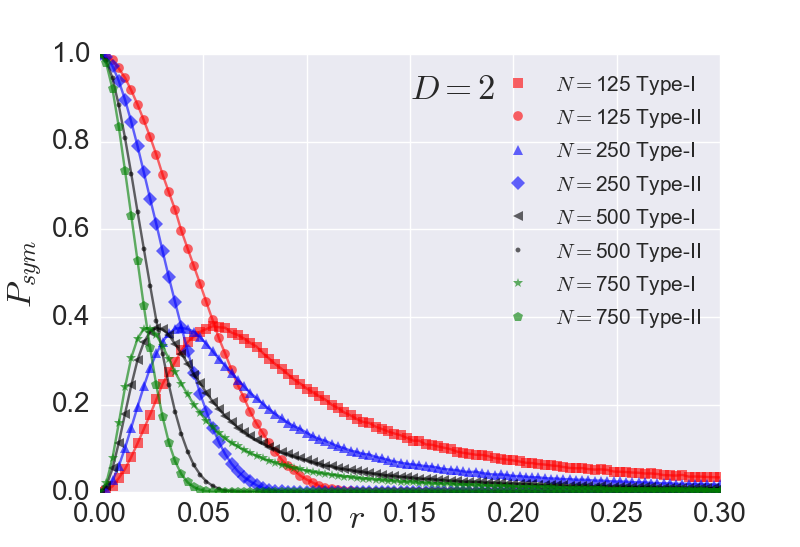} \includegraphics[width=6cm]{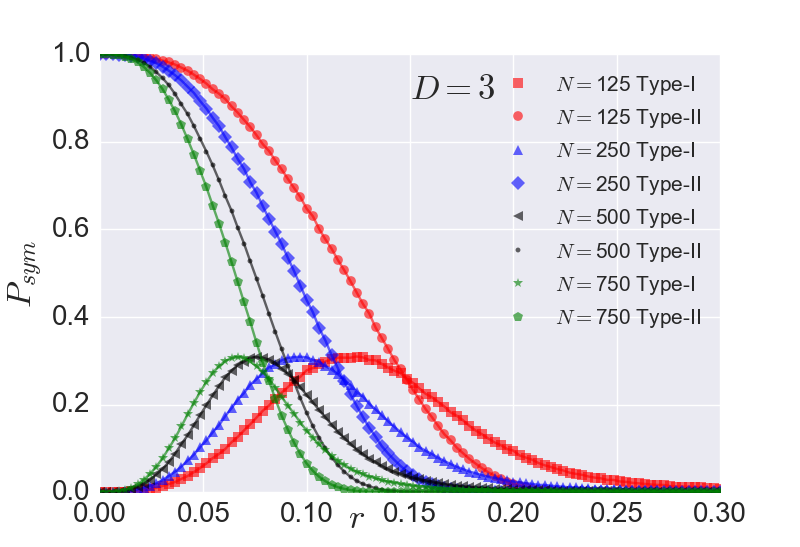}
\caption{Symmetry probability. Illustrated here is the probability $P_{sym}$ of a node being symmetric as calculated from ensembles of $10^3$ node RGGs as a function of connection radius $r$. We have calculated these as a function of dimension $D$ and intensity (number of nodes) $N$.}
\label{fig:symmProb}
\end{center}
\end{figure}

We calculated the probabilities of finding type-I and type-II symmetric nodes from ensembles of RGGs. The results are illustrated in figure \ref{fig:symmProb}. For one-dimensional RGGs the probability of type-I symmetric nodes quickly approaches a constant value as a function of connection radius whilst type-II quickly approaches zero. For two and three -dimensional RGGs we see entirely different behaviour. Most interestingly there is an optimal value of $r$ for finding type-I symmetric nodes.

\section{Dimension and limiting probabilities}
\label{sec:dimensions}

\begin{figure}[ht!]
\begin{center}
\includegraphics[width=6cm]{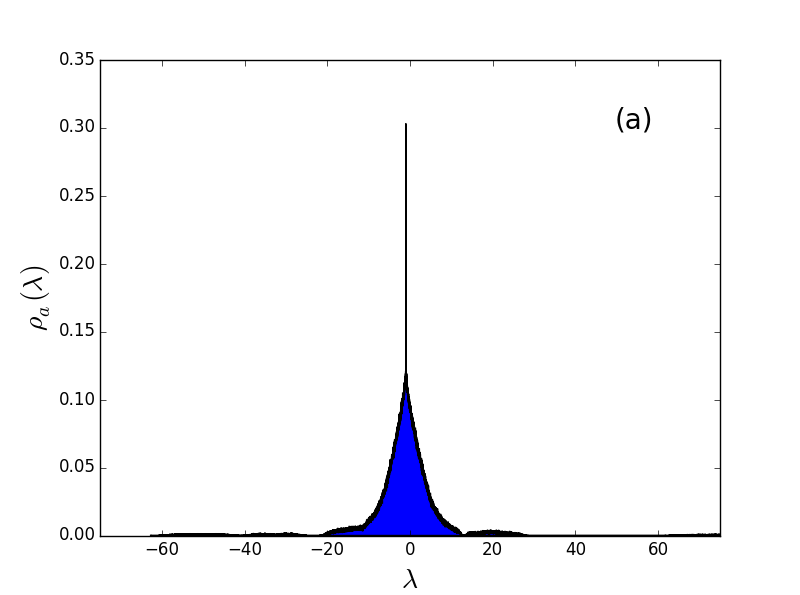} \includegraphics[width=6cm]{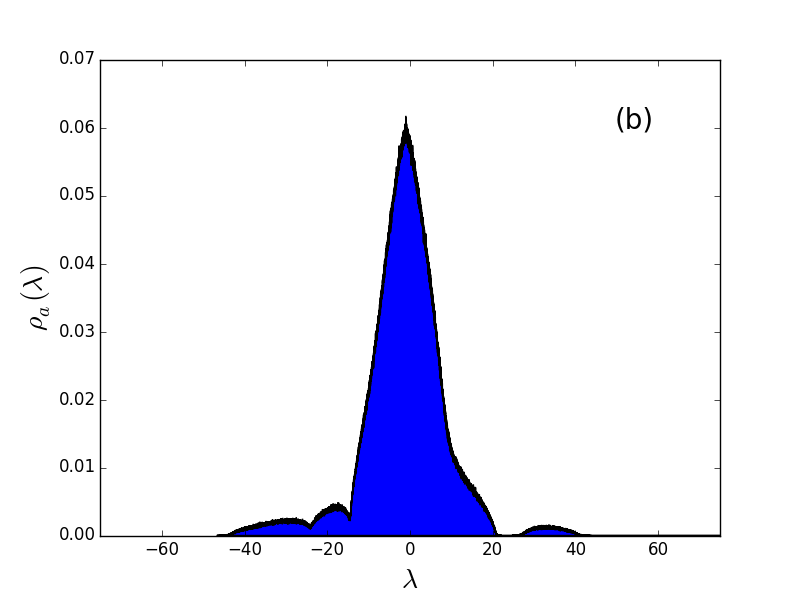}
\caption{3D RGG. Illustrated here is the adjacency spectral density of an ensemble of  $2D$ (left) and $3D$ RGGs. The RGGs consist of $1000$ nodes whilst the connection radius is $0.348569$ and $0.45$ for the $2D$ and $3D$ graphs respectively. This ensures that the connection probability is the same for both. }
\label{fig:3D}
\end{center}
\end{figure}

In order to look closer at the differences across dimension, we have looked at the adjacency spectral density of both $2D$ and $3D$ RGGs with similar connection probability. The results are illustrated in figure \ref{fig:3D} where we see a noticeable difference in that for the $3D$ RGGs the peak at $-1$ is far less prominent.

To understand this property we look at the probability that nearest neighbour vertices are symmetric (these are the most likely to be symmetric). We will analyse the minimum separation distance between nodes in a RGG and derive how it scales as the number of nodes increases.

We consider $N$ nodes $n_1, n_2,..., n_N$, uniformly distributed  on d-dimensional torus and choose an index set $I$ which consists of all pairs of nodes $I = \{ \alpha \subset \{ 1,2,...,N\}: | \alpha | = 2  \}$. For each $\alpha=\{i,j\}$ in the index set $I$ we let $X_{\alpha}$ be the indicator random variable of event  $|n_i -n_j| \leq x$. That is, the event that the nodes $n_i, n_j$ are separated by a distance less than $x$. Note that $ P(X_{\alpha}=1) = c_Dx^D = p_{\alpha}$, where $c_D$ is the volume of the unit ball in $D$ dimensions. Furthermore,  for each $\alpha \in I$ we choose a $B_{\alpha} \subset I$ with $\alpha \in B_{\alpha}$ such that $B_{\alpha}$ is a neighbourhood of dependence. That is  for $\beta \in B_{\alpha}, X_{\alpha}$ and $X_{\beta}$ are dependent. For each $\alpha$ we let 
\begin{equation}
B_{\alpha} = \{ \beta  \in I : \alpha \cap \beta \neq \emptyset\}.
\label{Eq:depend}
\end{equation}
Consider the sum 
\begin{equation}
W = \sum_{\alpha \in I} X_{\alpha},
\label{Eq:rvSum}
\end{equation}
and note that 
\begin{equation}
P(W=0) = P(s_{min} > x).
\label{Eq:SumProbEquivalence}
\end{equation}
where $s_{min}$ is the minimum distance of any pair of nodes in the RGG.
The expectation of $W$, is given by
\begin{equation}
\mathbb{E}(W) = {N \choose 2} c_Dx^D \\
\label{Eq:ExpW}
\end{equation}
We now apply the Chen-Stein method described in the Introduction to remove the dependency of the inter-node distances.  From Theorem $1$ in Ref.\cite{AGG} we have that
\begin{equation}
|P(W=0)-e^{-w}| \leq (b_1 + b_2 + b_3)\left(\frac{1-e^{-w}}{w}\right)
\label{Eq:Theorem1AGG}
\end{equation}
where $w = \mathbb{E}(W)$ and $b_1,b_2,b_3$ are constants given by
\begin{equation}
b_1 = \sum_{\alpha \in I} \sum_{\beta \in B_{\alpha}} p_{\alpha} p_{\beta}
\label{Eq:b1}
\end{equation}
\begin{equation}
b_2 = \sum_{\alpha \in I} \sum_{\alpha \neq \beta \in B_{\alpha}} p_{\alpha \beta}, \ \   p_{\alpha \beta} = \mathbb{E}(X_{\alpha}X_{\beta}).
\label{Eq:b2}
\end{equation}
\begin{equation}
b_3 = \sum_{\alpha \in I} \mathbb{E} | \mathbb{E} \left( X_{\alpha}-p_{\alpha} | \sigma(X_{\beta}: \beta \notin B_{\alpha})\right)|
\label{Eq:b3}
\end{equation}
From Eq.(\ref{Eq:depend}) we have 
\begin{eqnarray}\nonumber
                          b_1 &=& |I| |B_{\alpha}|(c_D x^D)^2\\\nonumber
                                &=& {N \choose 2} \left[  {N \choose 2} - {N-2 \choose 2} \right] (c_D x^D)^2\\
                                &=& (N^3 -5N^2/2 + 3N/2)(c_D x^D)^2.
\label{Eq:b1Eval}
\end{eqnarray}
Considering $b_2$ we note that the $X_{\alpha}$ are pairwise independent, this implies $\mathbb{E}(X_{\alpha}X_{\beta}) = \mathbb{E}(X_{\alpha}) \mathbb{E}(X_{\beta}) = (c_D x^D)^2$, therefore
\begin{eqnarray}\nonumber
                          b_2&=& |I| |B_{\alpha}-1|(c_D x^D)^2\\\nonumber
                                &=& {N \choose 2} \left[  {N \choose 2} - {N-2 \choose 2} -1  \right] (c_D x^D)^2\\
                                &=& (N^3 -3N^2 + 2N)(c_D x^D)^2.
\label{Eq:b2Eval}
\end{eqnarray}
Finally we note that $X_{\alpha}$ is independent of all $X_{\beta}$ where $\beta  \notin B_{\alpha}$, hence $b_3 =0$. Combining Eqs.(\ref{Eq:b1Eval},\ref{Eq:b2Eval}) via Eq.(\ref{Eq:Theorem1AGG}) we have the following
\begin{equation}
 |P(W=0) -  e^{-w}| \leq (4N-7)c_D x^D\left( 1-e^{-\frac{N^2-N}{2}c_Dx^D}\right).
\label{Eq:Inequality}
\end{equation}
We note that from Eq.(\ref{Eq:ExpW}) we have that $e^{-w}$ is of order one when $N^2x^D \sim 1$, that is when  $x \sim N^{-\frac{2}{D}}$. Furthermore in this regime from Eq.(\ref{Eq:Inequality})
\begin{equation}
 |P(W=0) -  e^{-w}|  \rightarrow 0, \ \ N \rightarrow \infty.
\label{Eq:limit}
\end{equation}
Hence we see that $P(W=0) = P(s_{min} > x)$ is order one when $x \sim N^{-\frac{2}{D}}$. From this we derive that $s_{min}$ scales like $N^{-\frac{2}{D}}$ for large $N$.

For $D=1$, two points separated by a distance $s$ have an excluded neighbourhood  of length $2s$. The probability that they are type-I symmetric with $s=s_{min}= C_1 N^{-2}$ for constant $C_1$ is then
\begin{eqnarray}
\pz(N(\mathcal{N}_{ex})=0) &=& \left(1- \frac{2C_1}{N^2}\right)^{N-2}\\
                           &\rightarrow& 1, N \rightarrow \infty.
\label{Eq:symm_prob_limit_1d}
\end{eqnarray}
So the closest nodes will be type-I symmetric in this limit. We obtain the same result in the thermodynamic limit, where the mean degree is kept constant with $r = C N^{-\frac{1}{D}}$.

For $D=2$, two points separated by a distance $s$ have an excluded neighbourhood equal to two times the area of a circle with radius $r$, minus four times the area of the circular segment with height $r-s/2$, that is
\begin{eqnarray}\nonumber
                        || \mathcal{N}_{ex}||&=&2\pi r^2-4r^2\cos^{-1} \left(  \frac{s}{2r} \right)  +2s\sqrt{r^2-\frac{s^2}{4}}\\
                                                             &=&4r^2 \sin^{-1}\left( \frac{s}{2r}\right)+2s\sqrt{r^2-\frac{s^2}{4}}
\label{Eq:2D_excludedneighbour}
\end{eqnarray}
Using $s=s_{min} = C_2 N^{-1}$ for some constant $C_2$ the probability of type-I symmetry in two dimensions is then
\begin{eqnarray}\nonumber
\pz(N(\mathcal{N}_{ex})=0)&=&\left(1-4r^2 \sin^{-1}\left(\frac{C_2}{2rN}\right)-\frac{2C_2}{N}\sqrt{r^2-\frac{C_2^2}{4N^2}} \right)^{N-2} \\
        &=& \left( 1 -\frac{2C_2}{N}\left(r+\sqrt{r^2-\frac{C_2^2}{4N^2}} \right)- \mathcal{O}(N^{-3})\right)^{N-2} \\
                         &\rightarrow& e^{-4C_2r}, N \rightarrow \infty
\label{Eq:Symmetry_probability_2D}
\end{eqnarray}
where we use $\sin^{-1}(x) = x + \mathcal{O}(x^3)$. In the thermodynamic limit with $r=CN^{-\frac{1}{2}}$
\begin{eqnarray}\nonumber 
\pz(N(\mathcal{N}_{ex})=0)&=& \left( 1 - \frac{4C^2}{N}\sin^{-1}\left( \frac{C_2}{2CN^{0.5}}\right)-\frac{2C_2}{N}\sqrt{\frac{C^2}{N}-\frac{C_2^2}{4N^2}} \right)^{N-2} \\\nonumber
                 &=& \left( 1 -\frac{2C_2C}{N^{\frac{3}{2}}}\left(1+\sqrt{1-\frac{C_2^2}{4C^2N^2}} \right)- \mathcal{O}\left(N^{-3/2}\right)\right)^{N-2} \\
                         &\rightarrow& 1, N \rightarrow \infty,
\label{Eq:Symmetry_probability_2D_tl}
\end{eqnarray}
this probability goes to one.

With $D=3$, two points separated by a distance $s$ have an excluded neighbourhood equal to two times the volume of a sphere minus four times the volume of the spherical cap with height $r-s/2$, that is
\begin{eqnarray}\nonumber
                          || \mathcal{N}_{ex}||&=&\frac{8\pi r^3}{3}-\frac{4\pi}{3}\left (r-\frac{s}{2}\right)^2 \left(2r+ \frac{s}{2}\right) \\ 
                                                             &=&2\pi sr^2-\frac{\pi s^3}{6}
\label{Eq:3D_excludedneighbour}
\end{eqnarray}
Using $s=s_{min} = C_3 N^{-\frac{2}{3}}$ for constant $C_3$,
\begin{equation}
                          || \mathcal{N}_{ex}||=2\pi r^2C_3 N^{-\frac{2}{3}}-\frac{\pi}{6}C_3 N^{-2}
\label{Eq:3D_excludedneighbour_lambda}
\end{equation}
Hence the probability of seeing no nodes in this excluded neighbourhood is
\begin{eqnarray}\nonumber
                \pz(N(\mathcal{N}_{ex})=0)&=& \left(1- 2\pi r^2C_3 N^{-\frac{2}{3}}+\frac{\pi}{6}C_3N^{-2}\right)^{N-2}\\
& \rightarrow 0&  N \rightarrow \infty.
\label{Eq:3D_symmetry_prob}
\end{eqnarray}

In the thermodynamic limit with $r=C N^{-\frac{1}{3}}$ 
\begin{eqnarray}\nonumber
                \pz(N(\mathcal{N}_{ex})=0)&=& \left(1- 2\pi C^2C_3 N^{-\frac{4}{3}}+\frac{\pi}{6}C_3N^{-2}\right)^{N-2}\\
& \rightarrow 1&  N \rightarrow \infty.
\label{Eq:3D_symmetry_prob}
\end{eqnarray}

The difference we see is that in one and two dimensions the probability that points separated by the minimal distance are symmetric approaches a constant value in the intensive limit of large $N$ whilst in three dimensions this probability goes to zero. In the thermodynamic limit this probability goes to one.

\section{Summary}
\label{sec:summary}

We have looked at the appearance of symmetric motifs in random geometric graphs. These subgraphs are of interest as they are prevalent in many real world networks and random geometric graphs but not in other random graph models. We looked at how the probability of finding symmetric nodes is dependent on the connection radius and the dimension of the random geometric graph. We found that in one dimensional random geometric graphs this probability is close to being independent of the connection radius and the density in that it quickly approaches a constant value as $r$ is increased. In two and three dimensions in contrast we found that there is a value of $r$ at which the probability attains a maximum value. In the thermodynamic limit we found that the closest nodes will be symmetric almost surely, irrespective of the dimension. Whilst in the intensive limit, in three dimensions this probability goes to zero.

Future work in this direction will look to analytically understand the numerical results we presented here on the probability of finding symmetric nodes as a function of connection radius. In addition, it will be possible to study how generalisations of the hard disc connection function affect the appearance of symmetric nodes.
\section*{Acknowledgment}
This work was supported by the EPSRC grant number EP/N002458/1 for the project Spatially Embedded Networks. The authors are grateful to an anonymous referee for correcting a few of the equations in Section 3.


%


\ifx\undefined\BySame
\newcommand{\BySame}{\leavevmode\rule[.5ex]{3em}{.5pt}\ }
\fi
\ifx\undefined\textsc
\newcommand{\textsc}[1]{{\sc #1}}
\newcommand{\emph}[1]{{\em #1\/}}
\let\tmpsmall\small
\renewcommand{\small}{\tmpsmall\sc}
\fi

\end{document}